\documentclass[twocolumn]{elsart}

\usepackage{natbib}

 \usepackage{graphicx}

\usepackage{amssymb}

\begin{document}

\begin{frontmatter}



\title{Soft X and Gamma ray emission from TeV sources observed
  with Swift and INTEGRAL}



\author[Rome]{Alessandra De Rosa}
\author[Rome]{Pietro Ubertini}
\author[Rome]{Angela Bazzano}
\author[Bologna]{Loredana Bassani}
\author[Bologna]{Raffaella Landi}
\author[Bologna]{Angela Malizia}
\author[Bologna]{John B. Stephen}
\author[] {and with behalf of IBIS Survey team  }

\address[Rome]{INAF/IASF, Rome}
\address[Bologna]{INAF/IASF, Bologna}

\begin{abstract}

The soft X-ray and soft gamma observations of the new discovered TeV sources by HESS
and MAGIC are crucial to discriminate between various emitting
scenarios and to fully understand their nature.
The INTEGRAL Observatory has regularly observed the entire galactic
plane during the first 1000 day in orbit providing a survey in the
20-100 keV range resulted in a soft gamma-ray sky populated with more
than 200 sources. 
In the case of HESS J1813-178 INTEGRAL found the hard X-ray
counterpart IGR J18135-1751 and Swift/XRT Telescope performed 
follow-up observations on this source.
Here we present the soft/hard X-ray spectral study.
We reported on the INTEGRAL observation of LS 5039, the first
microquasar that have been observed by HESS up to now.

\end{abstract}


\end{frontmatter}

\begin{figure*}
\begin{minipage}{5.cm}
 \includegraphics[width=5. true cm,height= 7 true cm,
 angle=-90]{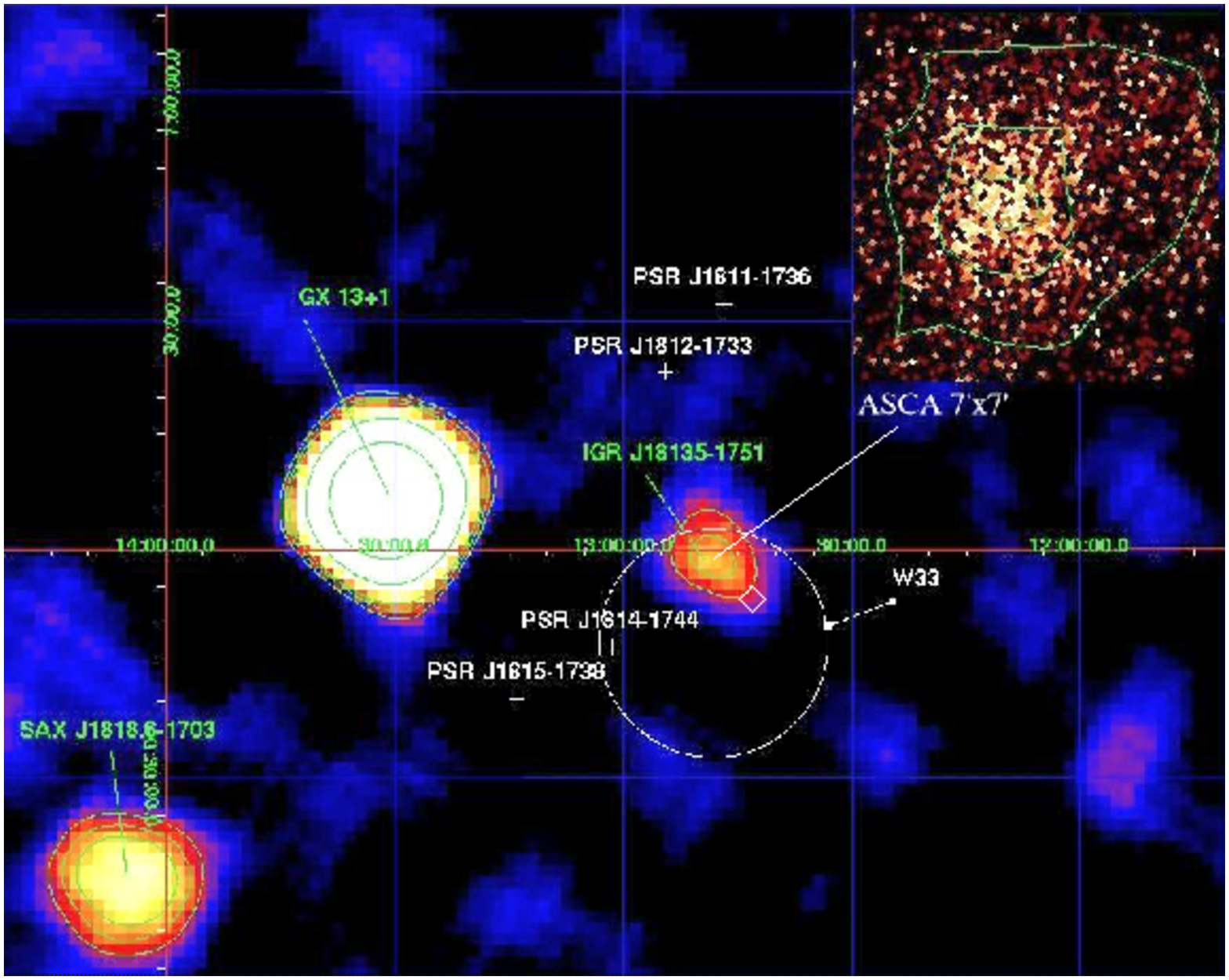}
\end{minipage}
\hspace{1.9cm}
\begin{minipage}{5.0cm}
\includegraphics[width=7 true cm,height= 5 true cm]{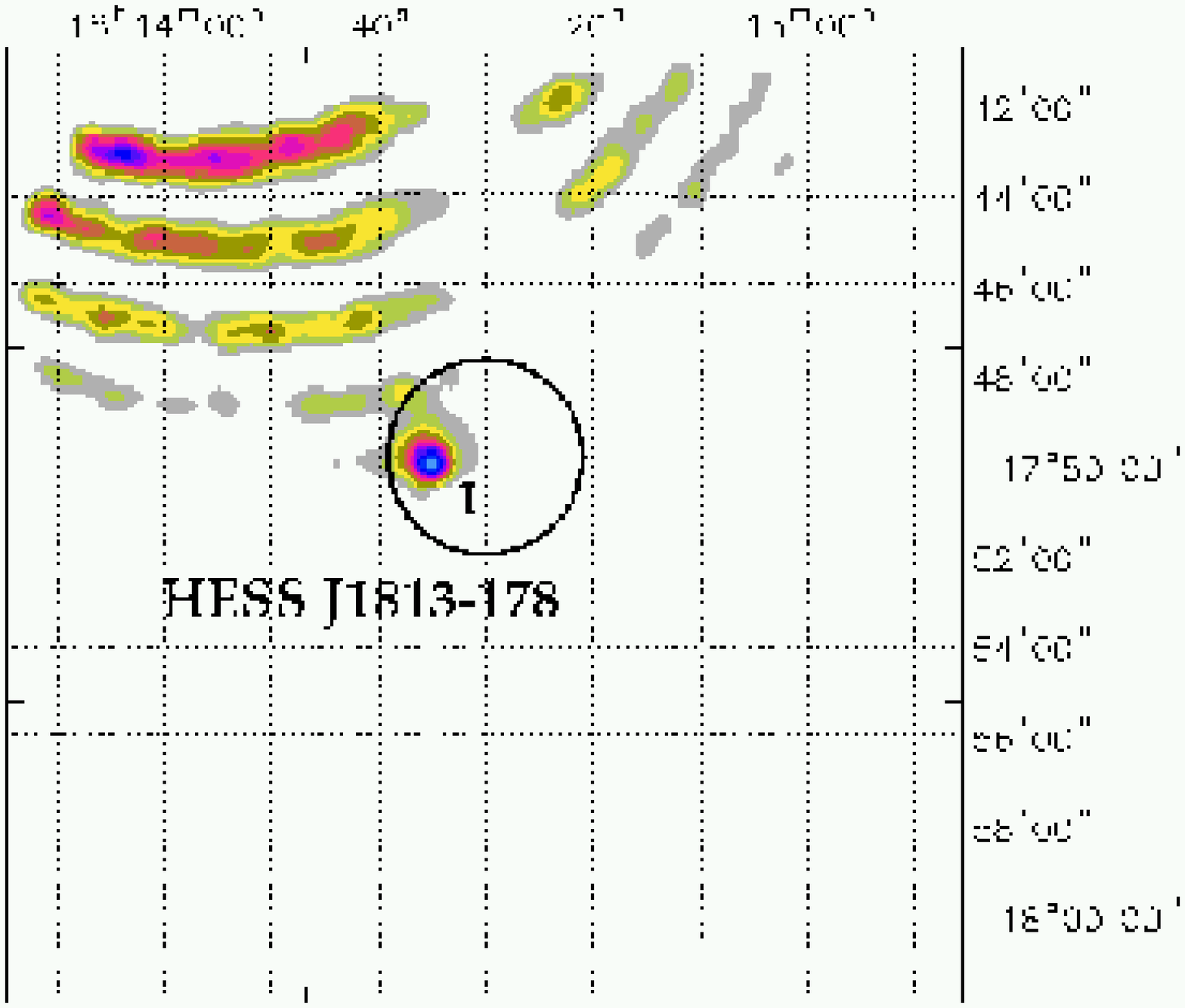}
\end{minipage}  
 \caption{Left panel. IGR J18135-1751: IBIS/ISGRI 20-40 keV
   map. Righ panel. The Swift/XTR image showing the extension of the
   HESS source}
  \label{igr}
\end{figure*}
%

\section{Introduction}
\label{intro}

HESS (High Energy Steroscopic System) collaboration have reported 
results of the
first sensitive TeV surveys of the inner part of our Galaxy \citep{aharonian},
revealing the existence of a population of high energy gamma-ray
objects, several of which previously unknown or not yet identified at
lower wavelength. Different types of galactic sources are known to be
cosmic particle accelerators and potential sources of high energy
gamma rays: isolated pulsars/pulsar wind nebulae (PWN), supernova
remnants (SNR), star forming regions, binary systems with a collapsed
object like a microquasar or a pulsar. Detection of X to gamma-ray
emission from these TeV sources is very important to discriminate
between  various emitting scenarios and, in turn, to fully understand
their nature. The IBIS gamma-ray imager on board INTEGRAL is a
powerful tool to search for their counterpart above 20 keV in view of
the arcmin  Point Source Location Accuracy associated to $~$ millicrab
sensitivity for exposure $>$1 Ms.
In addition SWIFT/XRT allow to obtain the accurate arc-second location
of the X-ray/radio counterpart of the
HESS sources. This is crucial to obtain secure
identification and to perform accurate optical/infrared follow-up.

We discuss here the INTEGRAL and Swift observation of IGR J18135-1751=HESS
J1813-178. The INTEGRAL observation of the microquasar LS 5039 
is also presented.


\section{IGR J18135-1751=HESS
  J1813-178: New Cosmic High Energy Accelerators from keV to TeV}
\label{}

IGR J18135-1751 was detected in the second IBIS/ISGRI survey
\cite{bird} with a significance exceeding
10$\sigma$ at R.A.(2000)= 18h 13m 27.12s and Dec(2000)=
-17$^{\circ}$ 50' 56'' (positional uncertainty of $<$3').
The averaged 20-100 keV flux was 2.1$\times$ 10$^{-11}$ erg cm$^{-2}$ s$^{-1}$.
HESS J1813-178 is one of the previously unknown sources
found in the HESS survey of the inner regions of the galactic
plane. This TeV source was also confirmed by MAGIC \citep{albert}.
 It is located at
R.A.(2000)=18h 13m 37.9s and Dec(2000)=-17$^{\circ}$ 50' 34''
(positional uncertainty of about 1-2'). The source does not
seem point -like although it is only slightly extended (3'), if
compared to the HESS point spread function. The statistical
significance of the TeV detection is around 9$\sigma$. The source is
fairly bright above 200 GeV with a flux of 12 $\times$ 10$^{-12}$
photons cm$^{-2}$ s$^{-1}$. No obvious counterparts were
found within the source extension.
At soft X-ray energies, we found
a possible counterpart in the ASCA archive data: AGPS273.4-17.8
at R.A.(2000) = 18h 13m 35.8s and Dec(2000) = -17$^{\circ}$ 49'
43.35'' with an associated uncertainty of 1'. 
In the X-ray band the source is fairly
bright showing a 2-10 keV flux (corrected for absorption) of 1.8
$\times$ 10$^{-11}$ erg cm$^{-2}$ s$^{-1}$ \cite {ubertini}.\\
In Figure \ref{igr} we show the IBIS/ISGRI 20-40 keV map
showing the location of
IGR J18135-1751 and relative significance contours: 6 (for the external one), 8, 10, 20 and 40
$\sigma$, compatible with a point source.  The extension of HESS J1813-178 as
well as the position of AGPS273.4-17.8 are both contained within
the internal IBIS/ISGRI contour. Also shown are the location (and
extension) of W33 and the 4 nearest radio pulsars (PSR J1814-1744,
PSR J1812-1733, PSR J1815-1738 and PSR J1811-1736).\\
The ASCA-SIS image is shown as an insert on the top right side of
the figure. The box covers an 8'x8' region centred on the ASCA
source position; the contour levels (1, 2 and 3 counts/pixel)
provide marginal evidence of extended emission.  GX13+1 and the
transient source SAX J1818.6-1703 also are visible in the image,
but contribute no contamination to region around IGR J18135-1751.
We find a bright NVSS (NRAO VLA Sky Survey,
\cite{con98}), radio source within the ASCA positional
uncertainty: NVSS J181334-174849 with coordinates R.A.(2000)=18h
13m 34.32s and Dec(2000)=-17$^{\circ}$ 48' 49.1''. It is contained within both
the IBIS/HESS circles, and possibly associated to the ASCA source.
The two observations of HESS J1813-178 performed with Swift/XRT
clearly show a point like object (RA: 18h 13m 34s.9s, DEC: -17$^\circ$
49' 53'') compatible with ASCA/INTEGRAL error boxes. 
In the left panel in Figure \ref{igr}, Swift/XRT 0.3-10 keV image of the region
surrounding HESS J1813-178 is shown. The extension of the TeV source 
is shown by the ellipse. The XRT source is almost 
at the center of the SNR shell G1282-0.02. Contamination from a very
bright source (GX 13+1) is also visible in the image. The XRT source
has a possible couterpart in a 2 MASS/DENIS object at RA: 18h 13m
35s.06s, DEC: -17$^\circ$ 49' 52.4''. The XRT spectrum is compatible with
that from ASCA but a factor of 2 lower. 
 \begin{figure}
\centering
\includegraphics[width=7 true cm,height= 6. true cm]{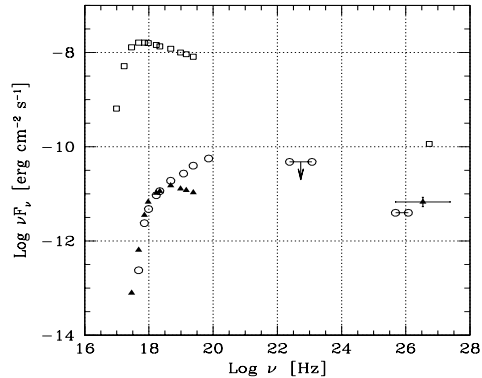} 
\caption{Spectral energy distribution of IGR J18135-1751=HESS
  J1813-178 (triangles) and
  AX J1838-0655/HESS J1837-069 (open circles). The SED of Crab nebula is also plotted
   (open squares).}
  \label{sed}
\end{figure}

HESS J1813-178 has a point like X-ray counterpart with a power law emission  from 2
to 100 keV and an associated radio counterpart. It is a non-thermal source,
possibly accelerating electrons and positrons which radiate through
synchrotron and inverse Compton mechanism. This is suggestive of
the presence of a PWN/SNR, as already found in most newly detected TeV
objects \cite{aharonian} that have been clearly associated with either
shell-type or plerion-type supernova remnants, like 
AX J1838-0655/HESS J1837-069 \citep{malizia}
The lack of strong X/Gamma variability is IGR J1813-178 as well as AX
J1838-0655, makes unlikely the scenario in which the TeV emission is
due to a binary system with a pulsar as compact object as observed in
HESS J1303-631/PSR B1259-63 \citep{aharonian2}.
In Figure \ref{sed} we plotted the SED of
both sources together with the SED of Crab nebula. The shape of the
curves is similar although with a quite different ratio X-ray to TeV
gamma rays. Also for PWNs, the ratio between the luminosity in X-ray
and that in radio is still an open issue: there
are sources of this type with X-ray luminosities similar to the Crab but
with  radio fluxes 2 to 3 order of magnitude weaker \citep{slane}.
Finally the observed X-ray, soft gamma and TeV luminosities 
(4, 3.4, (1.2-1.9) $\times 10^{34}$ erg/s for IGR J1813-178
and 10,32 and 4$\times 10^{34}$ erg/s for AX J1838-0655) are similar to the values observed
in the few HESS sources which have been clearly identified with PWNs
or shell type SNRs.

\section{Very high energy gamma rays from micro-quasar:LS 5039}
 
\begin{figure}
\centering
\includegraphics[width=7 true cm,height= 6. true cm]{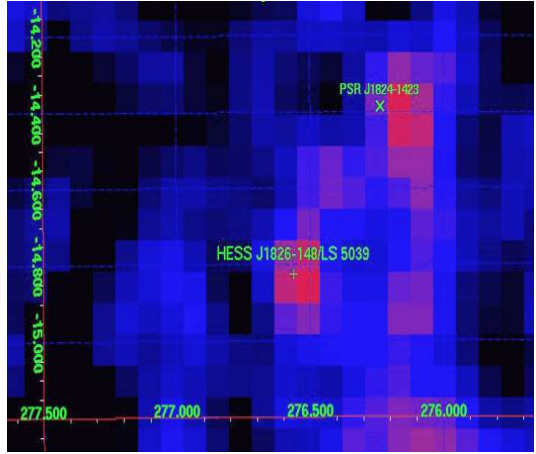} 
\caption{ IBIS/ISGRI 20-100 keV mosaic of ~900 Ks centered on LS
  5039. }
\label{ls5039} 
\end{figure}

The detection of microquasar in very high energy gamma rays
by HESS and MAGIC provide clear evidence that these objects are capable
of accelerating particles to TeV energies.
LS 5039 is a binary system composed by an O6.5 V-type donor star and a
compact object (probably a black hole) and it is the only microquasar
detected at TeV emission by HESS up to now \citep{aharonian3}.
In the Figure \ref{ls5039} we show the IBIS/ISGRI 20-100 keV mosaic of 
~900 Ks centered on LS 5039. 
The microquasar has been detected at more than 6$\sigma$ level.
The flux in 20--40 keV and 40--100 keV is (1.0$\pm$0.1)mCrab 
($(7.0\pm 0.7)\times 10^{-12}$ erg cm$^{-2}$ s$^{-1}$) and
(1.7$\pm$0.2)mCrab ($(1.6\pm 0.2)\times 10^{-11}$ erg cm$^{-2}$
s$^{-1}$. \cite{ubertini_elba}). These flux are in good agreement with the extrapolation of
RXTE/PCA data.
In the scenario of accretion-powered model, the high energy
emission is due to Inverse Compton (IC) with the electrons of the jet of a seed photon
field (produced by Synchrotron processes or by the optical-UV photons
of the donor star). The different energy losses processes,
synchrotron, Thomson IC or Kleyn-Nishina IC, produce
different ratio between X-ray and GeV-TeV emission.
A modeling of multiband data is then crucial, it was been recently
presented by \cite{goldoni}. Their major conclusion was that the soft
X-ray photons and the TeV photons are coming from different regions
while hard X-ray are due to synchrotron emission by the same
relativistic electrons that produce TeV radiation through IC.  \\
Simultaneous broad-band observation will be crucial to characterize
the relativistic electrons distribution, the seed photon field and the
magnetic field.

\section{Conclusions}

Ground based Cerenkov telescopes have discovered an increasing
population of TeV emitters, so far clustered in the Galaxy disk. 
The first sensitive survey by HESS have initially revealed several
objects, most of which without any obvious counterpart, though 
suggesting the existence of a new class of peculiar objects powered 
``exotic'' or non standard acceleration processes. More recently, the 
discovery of counterparts at radio, IR, X and soft-gamma Ray energies
has partially clarified the scenario suggesting that a large fraction 
of the TeV emitting sources have lower energies counterparts, often
associated with Pulsar Wind Nebulae. Nevertheless, different classes
of galactic sources can generate very high gamma-ray photons via
different acceleration processes. Among them SNR, Binary and isolated
pulsars, extragalactic objects. Finally at least two microquasars had
been clearly detected: LS 5039/RX J1826.2-145 \citep{aharonian3}
and LS I +61 303/V615 Cas \citep{albert2}.  Both of them are
associated with massive binary systems (O6.5 V and B0 Ve, coupled with a
NS or BH compact companion) with short period (4 and 26.5 day
respectively) and have counterparts in the IBIS/INTEGRAL soft
gamma-Ray domain. A plausible scenario is that the high energy
emission is due to particles accelerated in the moderately
relativistic double sided radio jets colliding via Inverse Compton
scattering with photons supplied by the star or by synchrotron
emission process, then boosting their energies in the hundreds of GeV to the TeV range.


We aknowledge financial contribution from contract ASI-INAF I/023/05/0.


\end{document}